# The study of high pressure structural stability of CeO$_2$ nanoparticles[*]

LIU Bo(刘波) LIU Ran(刘然) LI Quan-Jun(李全军) YAO Ming-Guang(姚明光) ZOU Bo(邹勃)

CUI Tian(崔田) LIU Bing-Bing(刘冰冰)[1]

State Key Laboratory of Superhard Materials, Jilin University, Changchun 130012, P. R. China

**Abstract:** In situ high pressure XRD diffraction and Raman spectroscopy have been performed on 12 nm CeO$_2$ nanoparticles. Surprisingly, under quasihydrostatic condition, 12 nm CeO$_2$ nanoparticles maintain the fluorite-type structure in the whole pressure range (0-51 GPa) during the experiments, much more stable than the bulk counterpart ($P_\text{T}\sim$31 GPa). In contrast, they experienced phase transition at pressure as low as 26 GPa under non-hydrostatic condition (adopting CsCl as pressure medium). Additionally, 32-36 nm CeO$_2$ nanoparticles exhibit an onset pressure of phase transition at 35GPa under quasihydrostatic condition, and this onset pressure is much lower than our result. Further analysis shows both the experimental condition (i.e., quasihydrostatic or non-hydrostatic) and grain size effect have a significant impact on the high pressure behaviors of CeO$_2$ nanomaterials.

## 1 Introduction

In recent decades, there has been a lot of interest in the high-pressure studies of nanostructured materials and their novel behaviors [1-3]. Previous studies indicate that phase transition pressures and processing of nanomaterials depend strongly on their grain size, shape, and structure [4-6].They have shown to have strong influence on critical pressures for phase transitions, phase transition routines, and even the amorphization processes [7, 8]. Recently, Gatta and his co-workers studied the elastic behavior of magnetite under high pressure using a mixture of methanol:ethanol:water = 16:3:1 as pressure-transmitting medium . They found that elastic behaviors of $Fe_3O_4$ were strongly affected by the experimental conditions. At about 9 GPa, the bulk modulus of $Fe_3O_4$ exhibited a significant increase, and the pressure medium was largely solidified. Further study shows that the anomalous behavior of $Fe_3O_4$ is due to the non-hydrostatic condition caused by the solidification of the pressure medium [9]. Quirke et al. studied the pressure dependence of the radial breathing mode (RBM) of carbon nanotubes. They found that the pressure dependence of the shift in vibrational modes of individual carbon nanotubes is strongly affected by the nature of the pressure medium as a result of adsorption at the nanotube surface. The adsorbate is treated as an elastic shell which couples with RBM of the nanotube via van der Waal interactions [10]. Hence, the hydrostatic or non-hydrostatic condition in high pressure studies plays a very important role in the behaviors of nanomaterials. Performing high pressure studies to explore how these conditions affect the nanomaterial behaviors is thus very important.

[*] Supported by the National Basic Research Program of China (2011CB808200), NSFC (10979001, 51025206, 51032001, 21073071, 11004075, 11004072, 11104105), and the Cheung Kong Scholars Programme of China.
[1] E-mail: liubb@jlu.edu.cn.

$CeO_2$ is an important material for numerous technological applications including catalyst, electrolyte, and solar cells due to its chemical stability, high oxygen storage capacity, etc [11]. Under ambient conditions, bulk $CeO_2$ crystallizes in cubic fluorite structure with (Fm3m) space group. It transforms into an orthorhombic $\alpha$-$PbCl_2$ structure at a pressure of 31 GPa [12]. When the crystal size is decreased to nanometer scale, $CeO_2$ shows different behaviors in phase transition pressures and routines. Rekhi and co-workers performed high pressure studies on 12 nm $CeO_2$ nanoparticles up to 36 GPa using in-situ high pressure Raman technique. They found that the transition from cubic to orthorhombic phase occurred at 26 GPa, significantly lower than that for the bulk materials [13]. However, in their study, CsCl was adopted as pressure medium, and thus apparent non-hydrostatic conditions existed in the whole process of their high pressure experiments [13]. According to recent researches, this non-hydrostatic condition was very likely to influence high pressure behaviors of $CeO_2$ nanomaterials. So far, to our knowledge, there has been no report on the study of the effect of hydrostatic or non-hydrostatic condition in the nano-$CeO_2$ system.

We have carried out high-pressure studies on 12 nm $CeO_2$ nanoparticles using *in situ* high-pressure X-ray diffraction under quasihydrostatic conditions. We found that under quasihydrostatic condition (using a mixture of methanol:ethanol = 4:1as the pressure medium ), 12 nm $CeO_2$ nanoparticles maintain the fluorite-type structure up to 51 GPa, much more stable than that of bulk $CeO_2$ ($P_T \sim$31GPa) [12]. And this finding is very different from previous studies [13, 14, 15]. Further analysis shows both the experimental condition (i.e., quasihydrostatic or non-hydrostatic) and the grain size effect have significant impact on the high pressure behaviors of $CeO_2$ nanomaterials.

## 2 Experimental details

12 nm $CeO_2$ nanoparticles were prepared by a simple solvothermal method using n-butanol as solvent. All reactants used were analytical grade without any further purification. 1mmol of $Ce(NO_3)_3 \cdot 6H_2O$ and 1mL of $NH_3 \cdot H_2O$ were loaded into a 50 mL Teflon-lined chamber which was filled with 40 mL of n-butanol. After being fully stirred to obtain a light brown solution, the autoclave was sealed and put inside an oven, which was maintained at 180 $^o$C for 15 h and then cooled to room temperature naturally. The resulting yellow brown precipitates were separated by centrifuging and washed with distilled water and ethanol several times, respectively. The final product was dried in air at 60 $^o$C for 24 h and collected for further characterization. A (HITACHI H-8100) transmission electron microscope (TEM) with accelerating voltage of 200 kV was employed to observe the morphology of the sample. X-ray powder diffraction (XRD) was used to characterize the product with Cu Kα radiation (λ = 0.15418 nm).When characterized by XRD, a scanning rate of 0.02° s$^{-1}$ was applied and the scanning range was 20°-90°. Raman spectrum was recorded on a Renishaw inVia Raman Microscope in the backscattering geometry using the 514.5 nm line of an argon ion laser, provided with a CCD detector system. Raman bands were analyzed by fitting the spectra to Lorentzian functions to determine the line shape parameters.

In situ high-pressure X-ray diffraction experiments were carried out up to 51 GPa using a synchrotron X-ray source ($\lambda = 0.6199$ Å) of the 4W2 High Pressure Station of Beijing Synchrotron Radiation Facility (BSRF). The diffraction data were collected using a MAR345 image plate. The two-dimensional X-ray diffraction (XRD) images were analyzed using the FIT2D software, yielding one-dimensional scattered intensity versus diffraction angle $2\theta$ patterns, which were then analyzed by GSAS + EXPGUI Rietveld package to obtain unit cell parameters. The average acquisition time was 150 s. High pressure Raman spectra were measured using a Raman microscope (Renishaw inVia) with 514.5 nm laser excitation up to 50 GPa. All high-pressure measurements were performed at room temperature. Pressures were generated in a diamond anvil cell with a culet size of 300 μm. The T301 stainless steel gasket was pre-indented by the anvils to an initial thickness of about 40 μm, and a center hole of 100 μm diameter was drilled as the sample chamber. A typical sample with a mixture of methanol and ethanol (4:1) as the pressure medium was loaded into the chamber, which provided a quasihydrostatic condition. Pressure was determined from the frequency shift of the ruby R1 fluorescence line.

## 3 Results and discussions

To characterize the morphology and structure of the $CeO_2$ nanoparticles, TEM and electron diffraction (ED) technique were carried out. Representative TEM micrographs of a typical product are shown in Fig. 1. These graphs reveal that all $CeO_2$ nanocrystals are monodispersive with an average size of 12 nm and a narrow diameter distribution. ED pattern (inset in Fig. 1) determines the polycrystalline nature of the product with a cubic fluorite structure.

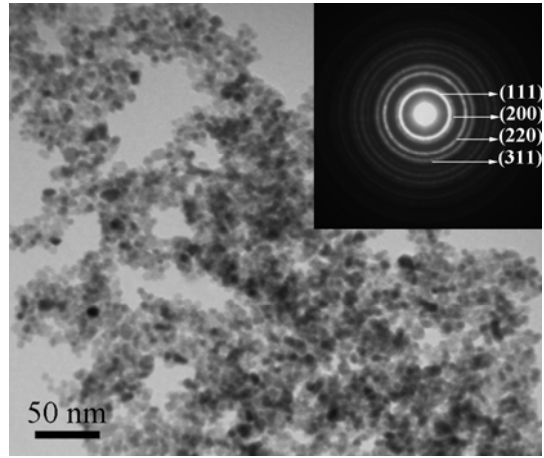

Fig. 1. The typical TEM and ED pattern of CeO$_2$ nanoparticles.

To determine the structure of the sample, XRD analysis was performed. Fig. 2 exhibits the XRD pattern of the CeO$_2$ nanoparticles. All detectable peaks in the pattern can be indexed to a pure cubic fluorite CeO$_2$ with a lattice constant a=5.411 (2) Å (JCPDS Card NO. 81-0792). According to the Debye-Scherrer formula, the strongest peak (111) at 2θ = 28.575°, the peak (200) at 2θ = 33.139°, the peak (220) at 2θ = 47.572° and the peak (311) at 2θ = 56.402° were used to calculate the average particle size of the CeO$_2$ nanoparticles, which was determined to be around 12 nm. These results are consistent with those obtained by TEM and ED analysis.

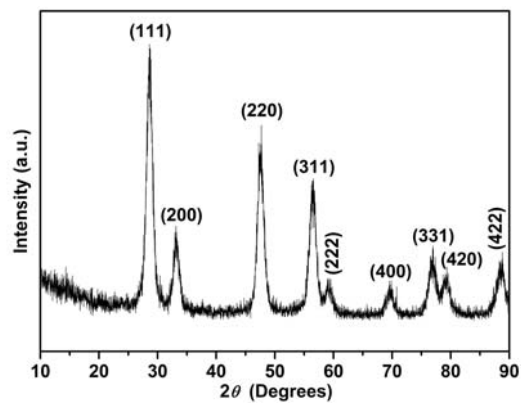

Fig. 2. The XRD pattern of CeO$_2$ nanoparticles.

Under ambint conditions, CeO$_2$ has a cubic fluorite type of structure and belongs to $O_{5h}$ (F m 3 m) space group. There is only one triply degenerate Raman active optical phonon ($F_{2g}$), which gives only one first order Raman line at about 465 cm$^{-1}$. In the second order Raman spectrum, with nine phonon branches, there are 45 possible tow phonon modes. The second order Raman peaks and their designations are 580, 660, 880, 1030 and 1160 cm$^{-1}$ for ωTO(X) + LA(X), ωR(X) + LA(X), ωLO + ωTO, 2ωR(X) and 2ωLO, respectively [16].

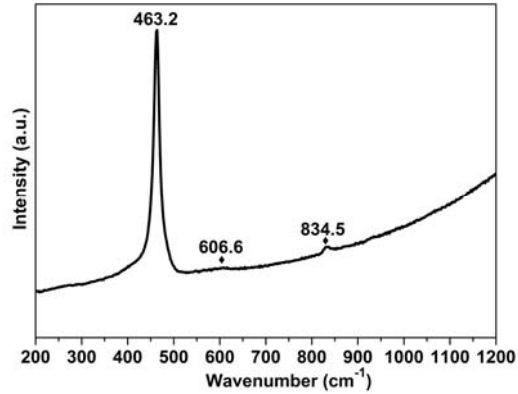

Fig. 3. The typical Raman spectrum of CeO$_2$ nanoparticles under quasihydrostatic condition using the mixture of methanol and ethanol (4:1) as pressure medium.

Fig. 3 shows the Raman spectrum of the CeO$_2$ nanoparticles with the fluorite phase within a range of 200-1200 cm$^{-1}$ under ambient condition. A first order Raman peak ($F_{2g}$) at 463.2 cm$^{-1}$ and a few second order Raman peaks at 606.6 and 834.5 cm$^{-1}$ were observed. These results are consistent with the above studies.

Fig. 4 shows in situ high-pressure X-ray diffraction patterns of the CeO$_2$ nanoparticles under quasihydrostatic condition using the mixture of methanol and ethanol (4:1) as pressure medium. It clearly shows that the 12 nm CeO$_2$ nanoparticles maintain the fluorite-type structure in the whole pressure range (0-51 GPa), much more stable than the bulk counterpart ($P_T \sim 31$ GPa) [12]. In contrast, previous in situ high pressure Raman and XRD studies showed that 12 nm CeO$_2$ nanoparticles exhibited less stabilty than bulk CeO$_2$, and had a phase transition pressure about 26 GPa [13, 14]. They used CsCl as the pressure medium and likely produced a non-hydrostatic condition. In order to illustrate our quasihydrostatic condition, we show the FWHM vs P in the whole pressure range. In Fig. 5, we can see that the XRD FWHMs almost remain constant in the whole pressure range, indicating the sample was under a quasihydrostatic condition. Thus, our comparative experiment shows that the quasihydrostatic or non-hydrostatic condition plays an important role in the high pressure behaviors of $CeO_2$ nanomaterials. Recently, S. Dogra et al. reported high pressure Raman study of $CeO_2$ nanoparticles using the mixture of methanol and ethanol (4:1) as pressure medium, and they found that 32-36 nm $CeO_2$ nanoparticles exhibited an onset pressure of phase transition at 35 GPa [15]. In contrast, our research show that $CeO_2$ nanoparticles are more stable with smaller particles size (~12 nm). The elevation of structural stability in the pressure-induced solid-solid phase transformation has been explained by the fact that smaller particle size leads to a higher surface energy between the phases involved in nanosized materials [17]. This result is in agreement with previous studies [18, 19]. Thus, our study shows that both the experimental conditions and the grain size effect have a strong impact on the high pressure stability of $CeO_2$ nanoparticles.

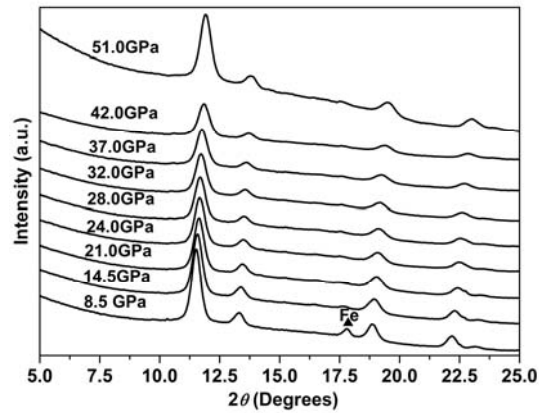

Fig. 4. In situ high-pressure X-ray diffraction patterns of 12 nm $CeO_2$ nanoparticles.

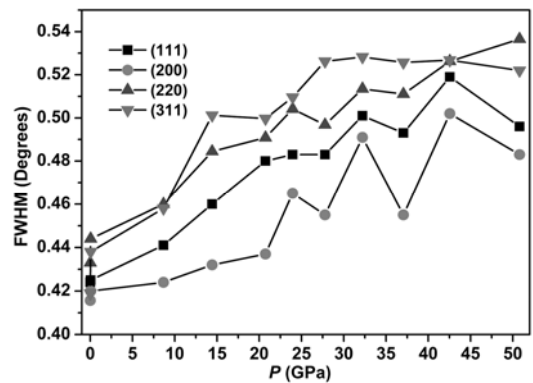

Fig. 5. The FWHM vs $P$ of 12 nm $CeO_2$ nanoparticles.

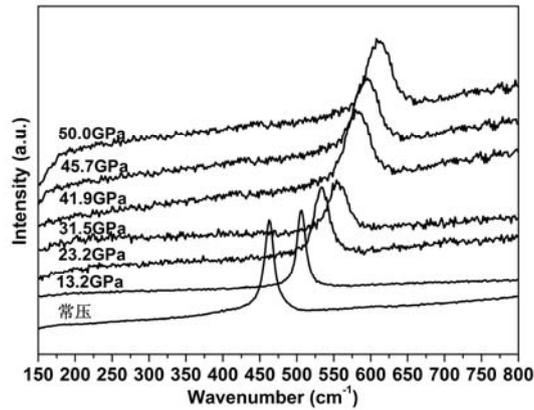

Fig. 6. In situ high-pressure Raman spectra of 12 nm $CeO_2$ nanoparticles.

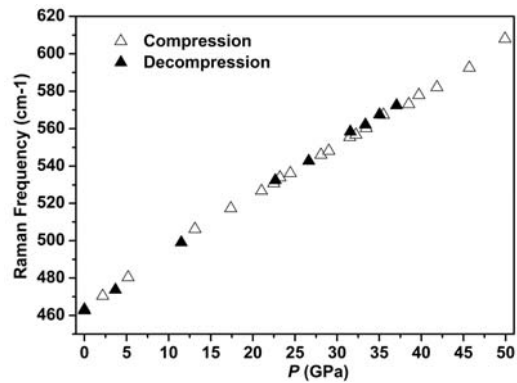

Fig. 7. The pressure dependence of the first-order Raman phonon frequency ($F_{2g}$) in the compression and decompression processes.

High pressure Raman spectra were collected for further studying the structural stability of 12 nm $CeO_2$ nanoparticles up to 50 GPa. Fig. 6 shows the in situ high-pressure Raman spectra of 12 nm $CeO_2$ nanoparticles. It is clearly observed in the whole pressure range, no other new peaks emerge, indicating that 12 nm $CeO_2$ nanoparticles maintain the cubic fluorite-type structure. The first-order fluorite peak ($F_{2g}$) shifts and slightly broadens with increasing pressure. It nevertheless still remains as the prominent peak up to 50 GPa. Fig. 7 exhibits the pressure dependence of Raman phonon frequency in the compression and decompression process from 0 to 50 GPa. During the compression to 50 GPa, we observed that the frequency of the first-order Raman peak increased with increasing pressure. Additionally, the change in this peak on decompression is similar to that in compression. This indicates there is no new phase in the decompression process. The high pressure Raman study is consistent with the high pressure XRD results.

## 4 Conclusions

We studied the effect of experimental conditions (i.e., hydrostatic or non-hydrostatic) on the phase stability of $CeO_2$ nanoparticles through in situ high-pressure X-ray diffraction technology and high-pressure Raman spectra. Surprisingly, under quasihydrostatic condition, 12 nm $CeO_2$ nanoparticles maintain the fluorite-type structure in the whole pressure domain (0-51 GPa), which is much more stable than the bulk counterpart ($P_T \sim 31$ GPa). In contrast, previous studies showed that 12 nm $CeO_2$ nanoparticles experienced phase transition at pressure as low as 26 GPa under non-hydrostatic condition (adopting CsCl as pressure medium), and 32-36 nm $CeO_2$ nanoparticles elevated the onset pressure up to 35 GPa while under quasihydrotatic cindition. Further analysis shows the experimental conditions (i.e., hydrostatic or non-hydrostatic) as well as the grain size effect have a significant impact on the high pressure behaviors of $CeO_2$ nanomaterials.

## 5 Acknowledgements

This work was supported by the National Basic Research Program of China (2011CB808200), NSFC (10979001, 51025206, 51032001, 21073071, 11004075, 11004072, 11104105), and the Cheung Kong Scholars Programme of China. We wish to thank Professor Keh-Jim Dunn for fruitful discussions and grammar revisions.